\documentclass[aps,prl,twocolumn,showpacs,superscriptaddress,groupedaddress]{revtex4}  
\usepackage{graphicx}
\usepackage{hyperref}
\usepackage{amsmath}
\begin{document}
\title{Further investigations on the Neutron Flux Generation in a Plasma Discharge Electrolytic Cell}
\author{R.~Faccini, A. Pilloni, A.D.~Polosa}
 \affiliation{Dipartimento di Fisica, Sapienza Universit\`a di Roma and INFN Sezione di Roma,\\ Piazzale Aldo Moro 2, I-00185 Roma, Italy}
 \author{ S. Loreti}
 \affiliation{ENEA, Centro Ricerche Casaccia, Via Anguillarese, 301, I-00123 Roma, Italy }
\author{ M. Angelone, E. Castagna, S. Lecci, A. Pietropaolo, M. Pillon, M. Sansovini, F. Sarto, V.~Violante}
 \affiliation{ENEA, Centro Ricerche di Frascati, Via E. Fermi 45, I-00044 Frascati, Italy }
 \author{R. Bedogni, A. Esposito}
 \affiliation{INFN Laboratori Nazionali di Frascati, Via E. Fermi 40, I-00044 Frascati, Italy }


\thispagestyle{empty}

\begin{abstract}
Our recent paper on the ``Search for Neutron Flux Generation in a Plasma Discharge Electrolytic Cell''~\cite{nostro} has as main goal the validation of the experiment in Ref.~\cite{Cirillo}. As a follow-up, Ref.~\cite{loro} moves a set of objections on our procedure and presents argumentations on why the experiments should not yield the same results. We collect here additional material and calculations that contribute to understanding the observed discrepancies. Furthermore we prove that the absence of signals from Indium activation detectors reported also for the experiment of Ref.~\cite{Cirillo} is a clear indication that neutron production does not occur.

\end{abstract}
 \maketitle

\section{Introduction}
Given the striking results obtained in Ref.~\cite{Cirillo} and the fact that some experimental aspects did not convince us, we set up to reproducing the experiment and published the results in Ref.~\cite{nostro}: we failed to reproduce the original results and we identified potential weaknesses in the measurements technique. In absence of undestanding the underlying physical processes, it is virtually impossible to reproduce exactly the original experiment, since any unavoidable small change in the setup can be pointed out as a cause of failure to reproduce the experiment. This of course speaks against the reproducibililty of the experiment, and suggests that only performing further experiments together could clarify the situation. On the contrary, considerations about the effectiveness of the neutron detection in an experiment have much more solid grounds, the biggest uncertainty being the energy spectrum of the generated neutrons.

A small note published on arXiv~\cite{loro} by the authors of Ref.~\cite{Cirillo} asks for further details to understand differences in the experimental setup and moves objections to our conclusions about the neutron detectors. Here we provide further material about our experiment and further argumentations on the neutron detectors. We follow the same structure of Ref.~\cite{loro} and respond point by point.

\begin{figure}[b!]
\centering
\includegraphics[width=0.7\columnwidth]{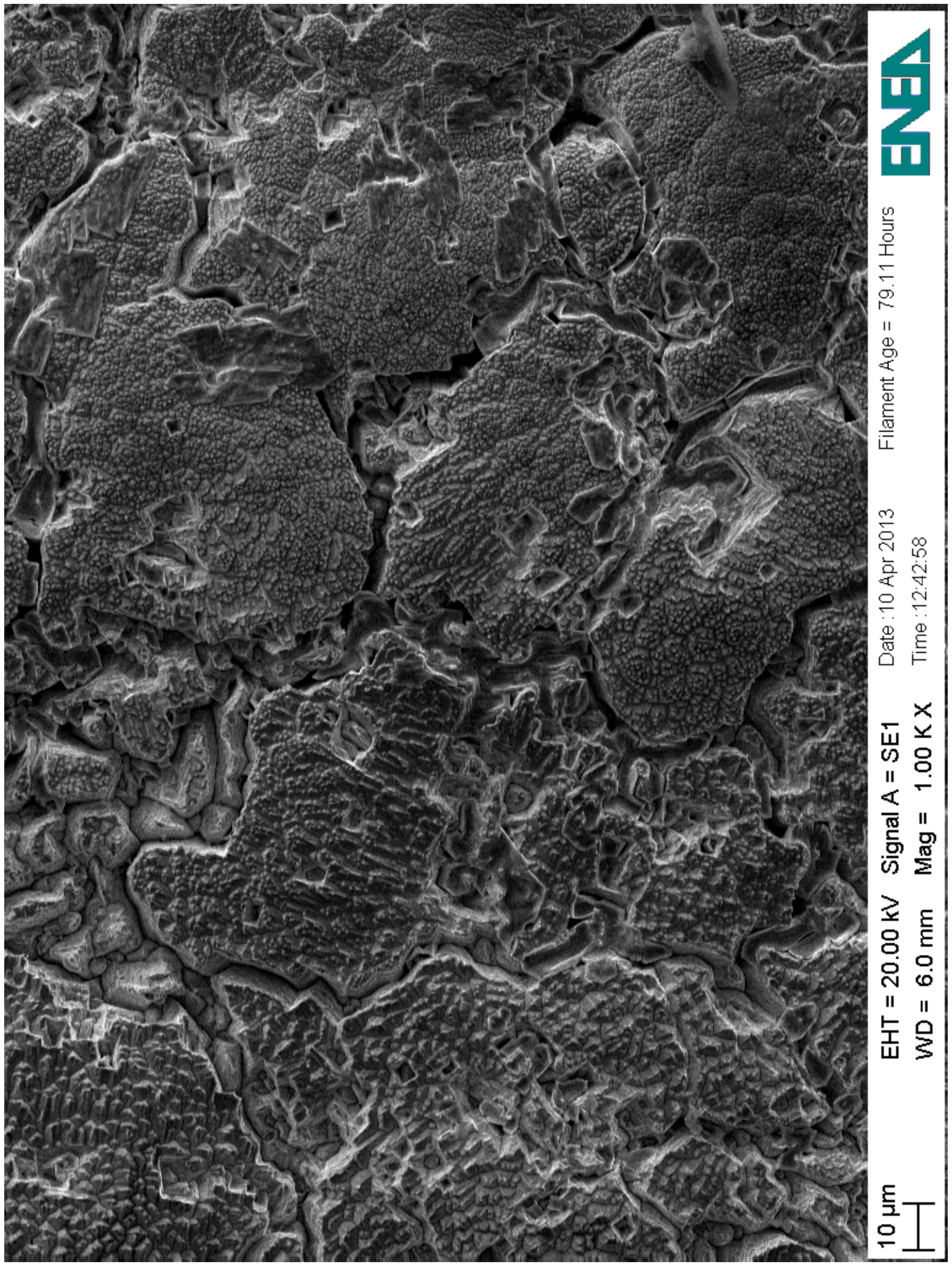}
\caption{Zoom of the cathodic surface.}
\label{fig:cathode}
\end{figure}

\section{Preparation and positioning of cathode and anode}
Ref.~\cite{loro} claims that the reaction of interest does not take place unless the ``cathode surface has cracks, with a sharp point but certainly not regular surfaces''. The required amount of roughness to have the alleged processes occur needs to be quantified in order to be reproducible. The cathode we used was not polished, as clearly shown in Fig.~\ref{fig:cathode}.

Also, the inner diameter of the quartz tube covering the diameter, present in both experiments, is considered a relevant parameter. Our was $3.3$~mm and the thickness was $1.3$~mm, but the authors do not provide a measurement of theirs to compare to.

Next, more details about the anode are requested: it is made of a platinum-plated aluminum  grid with cylindrical symmetry. This choice is typical in electrochemical applications, including the papers from Mizuno~\cite{mizuno}, because platinum is resistant to erosion and therefore avoids contamination of the solution. The choice of an iron structure with no cylindrical symmetry as used in~\cite{Cirillo} is instead unconventional and deserves justification.
\begin{figure*}[htb!]
\centering
\includegraphics[width=\textwidth]{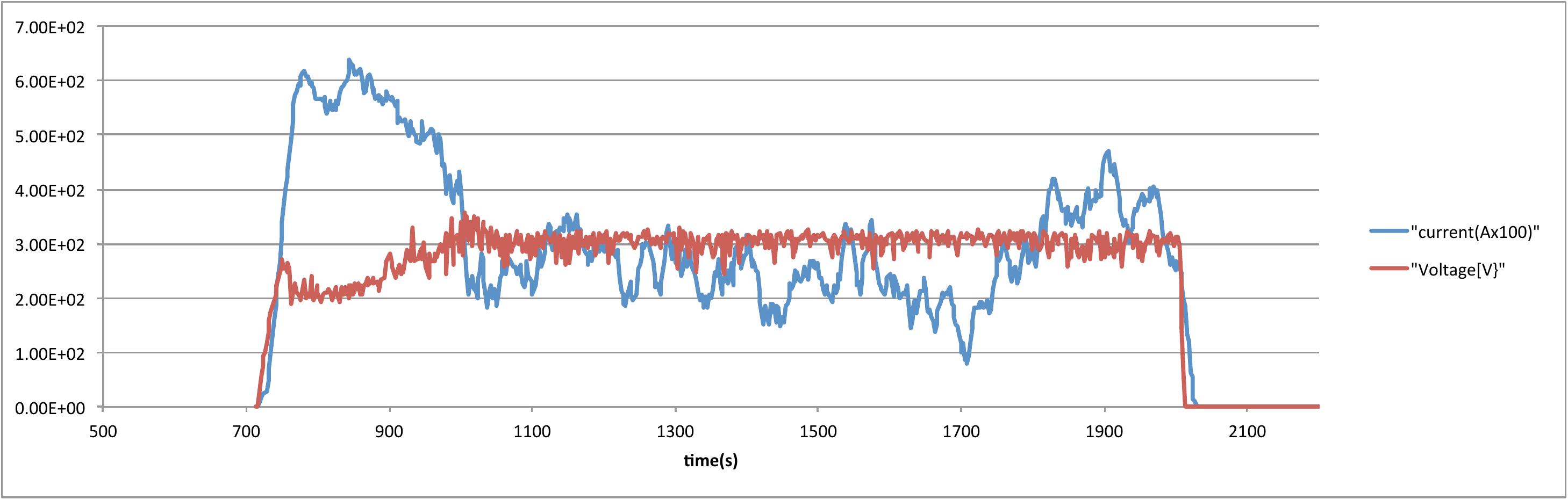}
\caption{Dependence with time of the voltage and the current in our cell for the Run3B run.}
\label{fig:IV}
\end{figure*}

Finally, it is noted that a significant difference is the presence in our experiment of a 4mm wall between anode and cathode. This was mutuated by the papers from Mizuno~\cite{mizuno}, has no impact on the plasma formation and allows to measure the gas flow separately in the anodic and cathodic region. Cirillo et al, should instead justify their decision to modify the cell, in particular if such change brings improvement in the neutron production, as claimed.


\section{Plasma Conditions}
The plasma conditions of our experiment are claimed to be unstable. This statement is based on the fact that the measurements on the CR-39 detectors were not correlated with the voltage. With this statement they clearly fail to get the core message of our paper: there are no neutrons and the ``measurements'' in the Run1 data are due to an experimental flaw. The lack of correlation between the small excess in the CR-39 detectors and the voltage supports such hypothesis and gives no indication about the plasma.

In any case, it is of interest to show the I-V curve and the stability with time. Fig.~\ref{fig:IV} shows the evolution with time of the voltage and the current measured for our most significant run, Run3B in our original paper. There is no significant different in stability with respect to the behaviour shown in Ref.~\cite{loro}. The voltage is in the range that is claimed to be appropriate and the current is 2-3 times larger than theirs.

Furthermore, the authors of Ref.~\cite{loro} judge the quality of the plasma in our experiment by the color of a picture. Several effects can of course affect the color of a digital image. In any case if this comforts them, the full video shows during most of the time the persistence of a white glow. 

\section{Neutron Detection}
The note~\cite{loro} provides the very important information, missing in the original paper, that Indium detectors were used and yielded a negative value. Their justification for such effect is that neutrons come in bursts and not from a continuous source. Because of the underlying physics mechanisms described in our original paper~\cite{nostro}, the time structure of the source cannot have any effect: while the activation of the Indium is instantaneous, the decay of the metastable state has a lifetime of 
54 minutes. This implies that there is no recovery effect in between the bursts. In other words,  effects at the sub-second level cannot be visible after folding with an exponential decay of almost an hour~\cite{pulsed}. As a further proof of this, such detectors were used to measure neutrons from plasma focus bursts of $100$-$200$~ns~\cite{plasmaIndium}.
We therefore stress again that the indium activation is  the most reliable detection strategy  in this context and that the absence of signals with Indium disks also in the case of Ref.~\cite{Cirillo} clearly indicates absence of neutrons in such experiment.

Next, Ref.~\cite{loro} claims their etching time for the CR-39 is different from ours and that the granularity of the Boron in our detectors makes our detector insensitive to Boron. Here the authors apparently failed to appreciate that the thick Boron configuration, were we see no signal, was considered with the  only purpose of understanding the behavior of the detectors used in Ref.~\cite{Cirillo}. The detectors for which we reported the results are covered with a $50\,\mu$m $^{10}$B~foils: additional boron would be a hindrance to the measurement of thermal neutrons, instead of favoring it. Any objection about the etching procedure is overcome by the calibration and background evaluation that was already reported in Ref.~\cite{nostro} and discussed further in the following.

The additional information contained in the note~\cite{loro} allows us to have a better picture of the discrepancies between their measurements and ours and here we summarize our findings.

\begin{figure}[b!]
\centering
\includegraphics[width=\columnwidth]{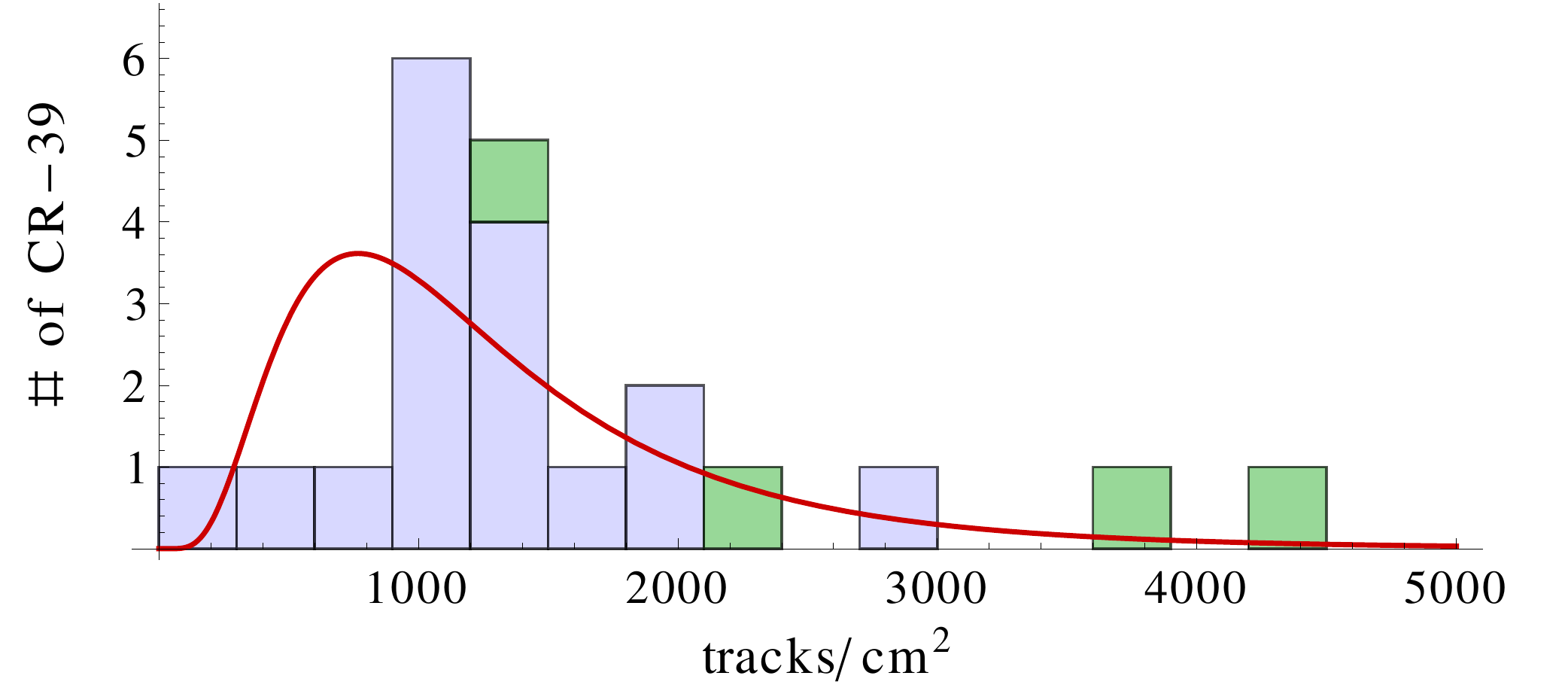}
\caption{Distribution of the background of the Cirillo {\it et al.} detectors as reverse engineered in this note. The fitted lognormal function is superimposed and the measurements on the exposed detectors are reported as well. }
\label{fig:resoCirillo}
\end{figure}

 The calibration procedures of the two experiments were the same: expose a sample of CR-39 detectors to the  INMRI-ENEA calibrated neutron flux and measured the correlation between the true number of impinging neutrons and the number of measured tracks.
The calibration constants obtained by us with the thin Boron film were 
\begin{align}
c_1&=  (6.9\pm0.3)\times 10^{-3} \text{ tracks}/\text{neutron}\\
c_{2,3}&=  (8.3\pm0.8)\times 10^{-3} \text{ tracks}/\text{neutron}
\end{align}
in Run1 and Run2/3 respectively. Conversely from the paper we could infer that (at least in the linear part of the curve) the calibration constant of Cirillo {\it et al.} is 
\begin{equation}
\label{calib_cirillo}
c_c= (1.00\pm 0.14)\times 10^{-4} \text{ tracks}/\text{neutron}
\end{equation}
This means that their detectors produce $\approx70$ times less tracks per impinging neutron than ours due to the excess of Boron, i.e. they are $\approx70$ times less sensitive. Our detectors with a thick Boron layer did not yield a signal because the layer was larger than theirs and because we used much shorted exposition times ($2$~min as opposed to $60$~min). With this considerations we believe that in both cases detectors were read properly and no further discussion is needed on the reading of the CR-39 or its calibration.

From this finding we can assess that the reduced sensitivity makes the detectors used in Ref.~\cite{Cirillo} even more sensitive to experimental uncertainties:
\begin{itemize}
\item in Ref.~\cite{nostro} we estimated an ambient contamination from fast neutrons or charged particles (i.e. insensitive to the presence of Boron) of $3.8$~tracks~cm$^{-2}$~day$^{-1}$. Applying the calibration constant~\ref{calib_cirillo} this implies that each day elapsed between the calibration and the day of the experiment creates a bias in the measurement of  $38000$~neutrons~cm$^{-2}$. For a 500~s run, this would imply a fake flux of $76$~neutrons~cm$^{-2}$~s$^{-1}$ for each day. The time interval in the case of Cirillo {\it et al.} has never been reported, but if it were 100 days, the ambient background would explain $\sim10\%$ of the observed signal. 

This is not critical, but it is an effect that needs to be accounted for. Furthermore, a large contribution of the contamination we measured can be due to radon whose concentration can depend strongly on the location of the experiment, both in terms of city and in terms of characteristics of the lab (for instance cellars have a particularly large contamination).
To have a perception of the possible impact, we considered that it has been estimated~\cite{radonmeas} that Radon causes on CR-39 detectors a rate of track density equal to 
\begin{gather}
\frac{dD}{dt} = k C \\
k \approx 60 \text{ tracks cm}^{-2}\text{ day}^{-1}/\left(\text{kBq m}^{-3}\right)
\end{gather}
where C is the contamination of radon (in~kBq~m$^{-3}$). From Ref.~\cite{radon} we estimated that in the city of Naples, where the laboratory of Cirillo is, the average radon contamination is  $\approx 230$~Bq~m$^{-3}$ and therefore the number of tracks caused on average on CR-39 detectors would be $\approx~13$~tracks~cm$^{-2}$~day$^{-1}$.  This implies that in 100 days the fake tracks would amount to more than 40\% of the largest observed signal.

Such guess-estimate depends on the readout system and does not take into account fluctuations of the radon contamination. It also depends on when the detectors were assembled and eventually on  the impact of the boron and of the cage. An estimate of such contamination cannot be done {\it a posteriori}, but needs to be part of the experiment.

\item  since we believe that a result is meaningful only if compared with the expected background distribution, we reverse engineered the latter. To this aim we extracted the resolution function, defined as the difference between the track density and the expected value in each detector. The former is reported for each detector in Fig.~2 of Ref.~\cite{loro}, while the latter is computed by multiplying the known flux of the INMRI neutron source by the time for which the detector was exposed (that can be found in slide 13 of  Ref.~\cite{CirilloDettaglio}) and the calibration constant in Eq.~\ref{calib_cirillo}.

The distribution of the background is given by the sum of the resolution function and the expected central value of the background $D_{bkg}=~624~$ tracks/cm$^2$, as estimated from the calibration curve. Fig.~\ref{fig:resoCirillo} shows the resulting estimated background distribution with a lognormal fit superimposed.

To estimate how likely are the measurements the track density measured on the exposed samples is also reported in the same figure.  

The two samples positioned outside the cell are consistent with the background while the other two detectors give a measurement that is marginally inconsistent with it. Tab.~\ref{tab:cirilloRes} details the confidence levels, as defined in Ref.~\cite{nostro} and the number of gaussian sigmas.

\item the calibration and the background studies were performed in a normal environment, while the measurements were under extreme temperature conditions. It is to be proven that the background fluctuations are not enhanced by this. It is easy to immagine that the high temperature could produce on CR-39 cracks that could be confused as tracks and the probability of the background to fluctuate to the measured values is even higher that what calculated in Tab.~\ref{tab:cirilloRes}.
\end{itemize}

\begin{table}[b!]
\begin{center}
\begin{tabular}{l|c|c|c}
Name & tracks$/$cm$^2$ & $\mathcal{P}\left(\mathcal{D}\right)$ & \# $\sigma$ \\ \hline
CR39 Sample 1 & 4450 & 98.9\% & 2.5 \\
CR39 Sample 2 & 3750 & 97.7\% & 2.3 \\
CR39 Sample 3-4 & 2113 & 85.4\% & 1.5 \\
CR39 Sample 5 & 1302 & 60.2\% & 0.85 \\
\end{tabular}
\end{center}
\caption{\small{Measured track density ($D$), probability of consistency with the background (C.L.) and number of gaussian sigmas of deviation from the background, for the samples described in Ref.~\cite{loro}}}
\label{tab:cirilloRes}
\end{table}

\section{Conclusions}
In absence of any understanding of the underlying physics and in lack of details, it is clearly impossible to be sure that the experimental conditions of Ref.~\cite{Cirillo} are reproduced, regardless of our efforts. The most reasonable test would be to exploit the competences in neutron detection of our group to perform the measurements during an experiment of Cirillo {\it et al.}
In any case there are unjustified deviations of the cell in the Cirillo {\it et al.} experiment from the original Mizuno cell, namely the choice of the anode, asymmetric and made of iron and  the absence of the separation between the anodic and cathodic region.

In any case, there are no physically sound arguments against the use of indium detectors in such context. The fact that in the Cirillo {\it et al.} experiment they yielded no signal is an evidence against the production of  neutrons in it.

Finally, the CR-39 detectors soaked in boron have a reduced sensitivity to neutrons and the calibration constant published in Ref.~\cite{Cirillo} confirms it. This makes them sensitive to background fluctuations, in particular given the extreme and unusual temperature conditions in which they are operated. We reverse engineered their resolution function and estimated that all signals showed where less than 3$\sigma$ significant without even taking into account the effect of operating in high temperature water.

Since these considerations should be done during the experiment and not extrapolated from the results, we believe that the Cirillo {\it et al.} experiment would have a sounder basis if:
\begin{itemize}
\item background were appropriately studied, by showing the distributions and by paying attention that the background samples are developed at the same time as the exposed ones.
\item the background is estimated under the same conditions of the exposed ones, i.e. in hot water (of course with no plasma involved).
\end{itemize}


\begin{thebibliography}{100}
\bibitem{nostro}
R.~Faccini, A.~Pilloni, A.~D.~Polosa, M.~Angelone, E.~Castagna, S.~Lecci, A.~Pietropaolo and M.~Pillon {\it et al.},
  \href{http://arxiv.org/abs/1310.4749}{arXiv:1310.4749 [physics.ins-det]}.
\bibitem{Cirillo}
D.~Cirillo, R.~Germano, V.~Tontodonato, A.~Widom, Y.N.~Srivastava, E.~Del~Giudice, and G. Vitiello, Key Engineering Materials 495, 104 (2012).  
 \bibitem{loro}
 A.~Widom, J.~Swain and Y.~N.~Srivastava,
  \href{http://arxiv.org/abs/1311.2447}{arXiv:1311.2447 [physics.ins-det]}.
  \bibitem{mizuno} T.~Mizuno, T.~Ohmori, and T.~Akimoto, ``Generation of Heat and Products During Plasma Electrolysis'' in Tenth International Conference on Cold Fusion,  2003, Cambridge.
 
  \bibitem{pulsed}D.~Antonini, P.~Moioli, E.~Pedretti, R.~Scaf\`e and A.~Visentin, ``Measurements of neutron activation cross sections by generalized intermittent irradiation", Nucl. Ins. Meth. 151 (1978) 567.
 \bibitem{plasmaIndium} 
 B.~Bienkowska, L.~Karpinski, M.~Paduch, M.~Scholtz, K.~Pytel, R.~Prokopowicz, and A.~Szydlowski, ``Measurements of neutron yield from PF-1000 device by activation method", Czech. J. Phys., 56 (2006) B 377; J.M.~Koh, R.S.~Rawat, A.~Patran, T.~Zhang, D.~Wong,
S.V.~Springham, T.L.~Tan, S.~Lee and P.~Lee, ``Optimization of the high pressure operation regime for enhanced neutron yield in a plasma focus device", Plasma Sources Sci. Technol. 14 (2005) 12.
 \bibitem{CirilloDettaglio}
  \href{http://www.22passi.it/pirelli/D_Cirillo_ISPRA_12_2012_ufficiale.pdf}{\tt http://www.22passi.it/pirelli/\\D\_Cirillo\_ISPRA\_12\_2012\_ufficiale.pdf}

\bibitem{radonmeas}
Matiullah, ``Determination of the calibration factor for CR-39 based indoor
radon detector", J. Radioanal. Nucl. Chem. (2013) 298:369; G.~Sciocchetti {\it et al.}, ``A new passive Radon-Thoron discriminative measurement system", Radiation Protection Dosimetry (2010), Vol. 141, No. 4, pp. 462.
 \bibitem{radon}
 \href{http://www.a2c.it/Radon-Rn-222/banca-dati-misurazioni-radon.html}{\tt http://www.a2c.it/Radon-Rn-222/\\banca-dati-misurazioni-radon.html}

\end{thebibliography}
\end{document}